\documentclass[
    ,final            % use final for the camera ready runs
  ]
  {aipproc}

\layoutstyle{6x9}

\begin{document}

\title{Measurement of J/$\psi$ production at the LHC with the ALICE experiment}

\classification{25.75.-q, 25.75.Cj, 14.65.-q, 13.85.-t
%%<Replace this text with PACS numbers; choose from this list:             \texttt{http://www.aip..org/pacs/index.html}>
}

\keywords      {J/$\psi$, LHC}

\author{M. Gagliardi, for the ALICE Collaboration}{
  address={INFN Sezione di Torino, V. Giuria 1 10125 Torino - Martino.Gagliardi@cern.ch}
}

%\author{the ALICE collaboration}

%\author{<author2>}{ address={<common address for author2 and author3>}}

%\author{<author3>}{address={<common address for author2 and author3>},altaddress={<author1 address>} % additional visiting address}

\begin{abstract}
 ALICE (A Large Ion Collider Experiment) aims at studying the behaviour of nuclear matter at high energy densities and the transition to the Quark Gluon Plasma (QGP), expected to occur in ultra-relativistic heavy ion collisions.
 Quarkonia production measurements in both \hbox{Pb-Pb} and p-p collisions play a crucial role in the ALICE physics program.
   Quarkonium detection is possible in ALICE at both forward (in the dimuon channel) and mid-rapidity (in the dielectron channel).
  In 2010, the Large Hadron Collider has provided p-p collisions at $\sqrt{s}$=7~TeV and Pb-Pb collisions at $\sqrt{s_{NN}}$=2.76~TeV. 
   The ALICE results on J/$\psi$ production in p-p collisions are presented\footnote{Preliminary results were presented at this conference. However, since final results are now available, obtained to a large extent with the same analysis strategy, the latter are reported in these proceedings, in agreement with the Organisers.}, along with the status of the Pb-Pb analysis.
\end{abstract}

\maketitle

%%%%%%%%%%%%%%%%%%%%%%%%%%%%%%%%%%%%%%%%%%%%
%% MAINMATTER
%%%%%%%%%%%%%%%%%%%%%%%%%%%%%%%%%%%%%%%%%%%%

\section{Introduction}

The ALICE~\cite{alice_JINST} experiment at the Large Hadron Collider~\cite{lhc_JINST} was designed to study strongly interacting matter in ultra-relativistic heavy-ion collisions, at energy densities up to two orders of magnitude larger than that of ordinary nuclear matter. 
Under these conditions, finite temperature QCD calculations on the lattice~\cite{karsch} predict a transition to a deconfined state of matter known as Quark-Gluon Plasma (QGP).
Heavy flavour particle production is modified by the presence of the medium. 
%Heavy flavour particles are sensitive to the properties of the medium formed in heavy ion collisions.
%For the J/$\psi$, possible scenarios range from suppression by colour screening~\cite{satz} to final-state regeneration due to the recombination of initially uncorrelated $c$ and $\overline{c}$ quarks~\cite{recomb1}. 
In particular, quarkonium production suppression by colour screening was one of the first proposed signatures for QGP formation~\cite{satz}; charmonium regeneration due to the recombination of initially uncorrelated $c$ and $\overline{c}$ quarks may occur at LHC energies~\cite{recomb1}. A detailed description of the physics motivations for charmonium measurements in heavy-ion  collisions can be found in ~\cite{PPR}. 
Quarkonium production measurement in \hbox{p-p} collisions is crucial as a reference for Pb-Pb data; moreover, it is expected to provide insights on the J/$\psi$ production mechanism.

\section{J/$\psi$ detection in ALICE}

A complete description of the ALICE experiment can be found in~\cite{alice_JINST}. 

The J/$\psi$ measurement at mid-rapidity is performed by detecting its dielectron decays with the ALICE central detectors (-0.9<$\eta$<0.9). This analysis is based on the Inner Tracking System (ITS) for vertexing and tracking and the Time Projection Chamber\cite{tpc} (TPC) for tracking and particle identification. Other detectors such as the Transition Radiation Detector, the Time of Flight and the Electromagnetic Calorimeter, not used in this analysis, will greatly improve the ALICE electron identification capabilities in the future. 

At forward rapidity, J/$\psi$ production is measured in the dimuon channel with a dedicated muon spectrometer (-4<$\eta$<-2.5). The muon spectrometer is equipped with a dipole magnet, a set of absorbers, five tracking stations %based on multiwire proportional chambers with pad readout 
and a dedicated trigger system.% with programmable cut on the muon $p_{\rm T}$.% ($\simeq$0.5~GeV/c in 2010)

The mid-rapidity analysis is based on minimum-bias events, triggered by the V0 scintillators and the  two innermost layers of the ITS (Silicon Pixel Detector, SPD). The forward rapidity analysis is based on muon triggered events.
Luminosity normalisation is performed via a reference cross section, measured in van der Meer scans with the V0 detector~\cite{Ken}.

The results presented here are for inclusive J/$\psi$ production: promptly produced J/$\psi$s are not separated from those originating from B meson decays. 
The dependence of acceptance on the unknown J/$\psi$ polarisation is a major source of systematic uncertainty (up to~\hbox{$\simeq$30\%}) for both the dielectron and the dimuon channel. Here, we assume no polarisation and quote separately the polarisation-related uncertainty.

\section{J/$\psi$ $\rightarrow$ ${\rm e^+e^-}$ at mid-rapidity}\label{sec:jpsiee}

\begin{figure}[!hbtp]
   \centering
  
           \includegraphics[width=0.45\textwidth,height=0.23\textheight]{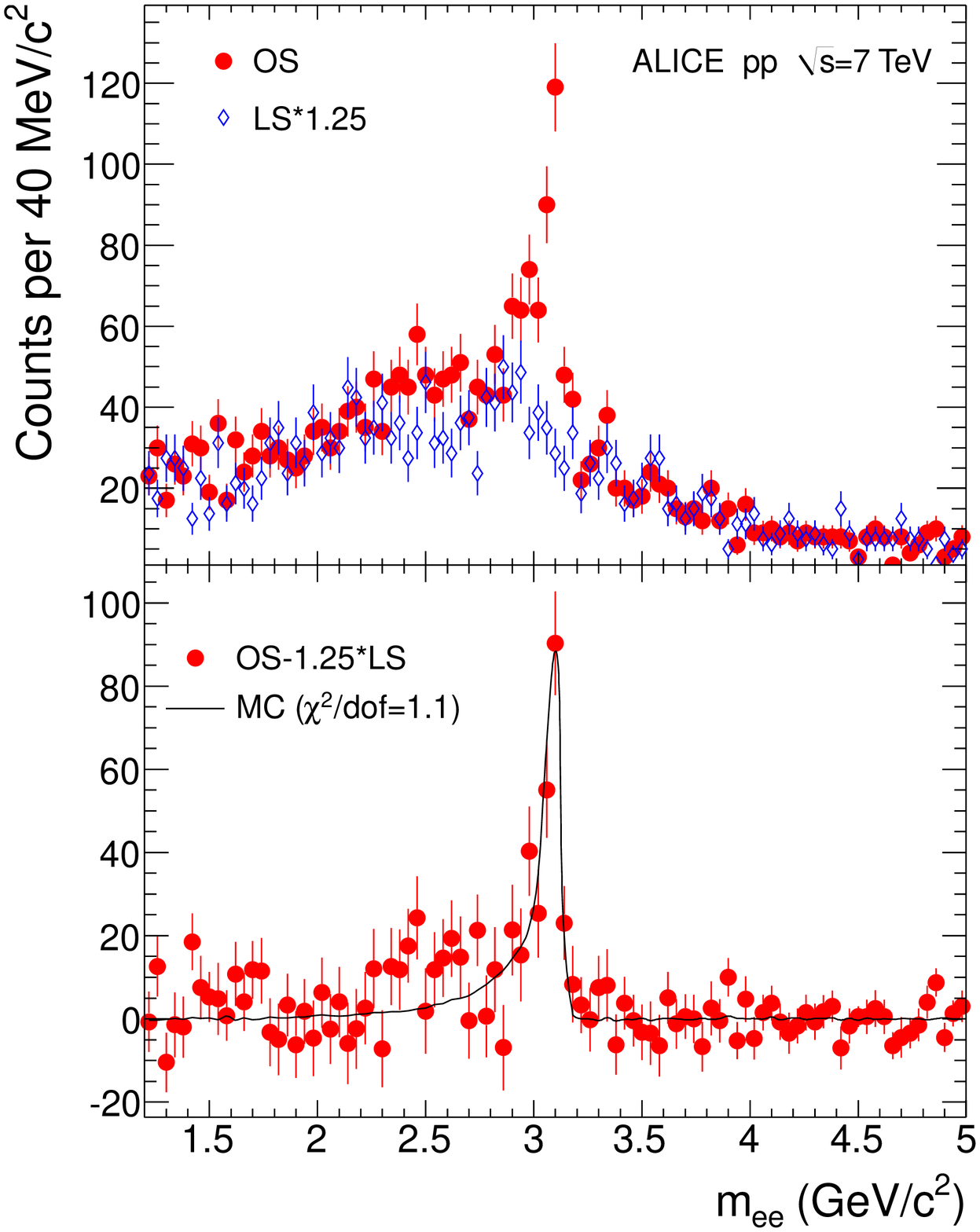}
	   \includegraphics[width=0.45\textwidth,height=0.23\textheight]{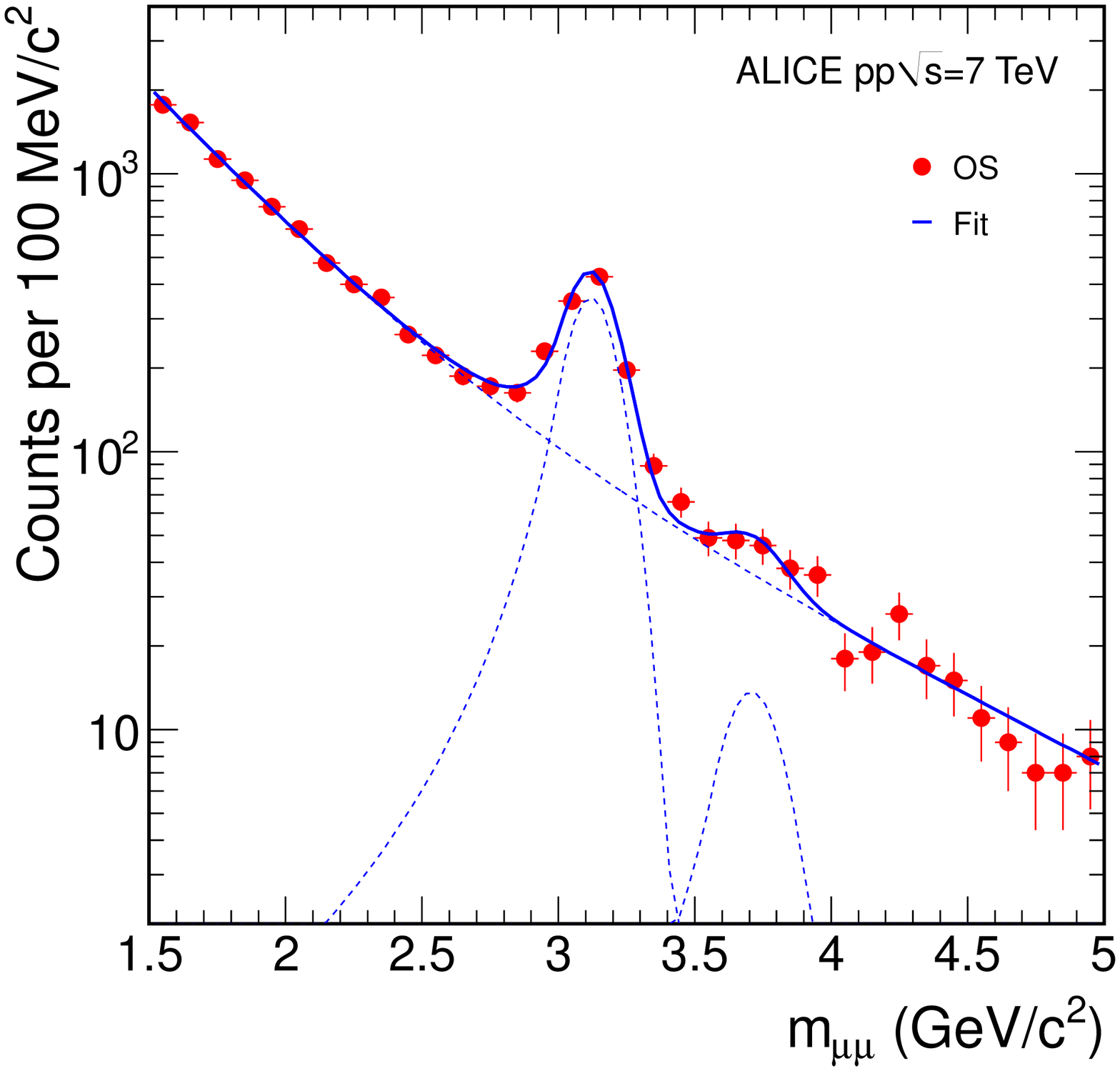}
           \label{fig:signal}
   \caption{Left: invariant mass distributions for like-sign (LS) and opposite-sign (OS) electron pairs (top panel); difference of the two distributions, compared to a Monte Carlo (MC) signal shape (bottom panel). Right: invariant mass distribution for opposite-sign muon pairs.  The fitted J/$\psi$ and $\psi$ (2S) contributions, as well as the background, are also shown.}

\end{figure}

\begin{figure}[!hbtp]
   \centering
  
           \includegraphics[width=0.45\textwidth,height=0.23\textheight]{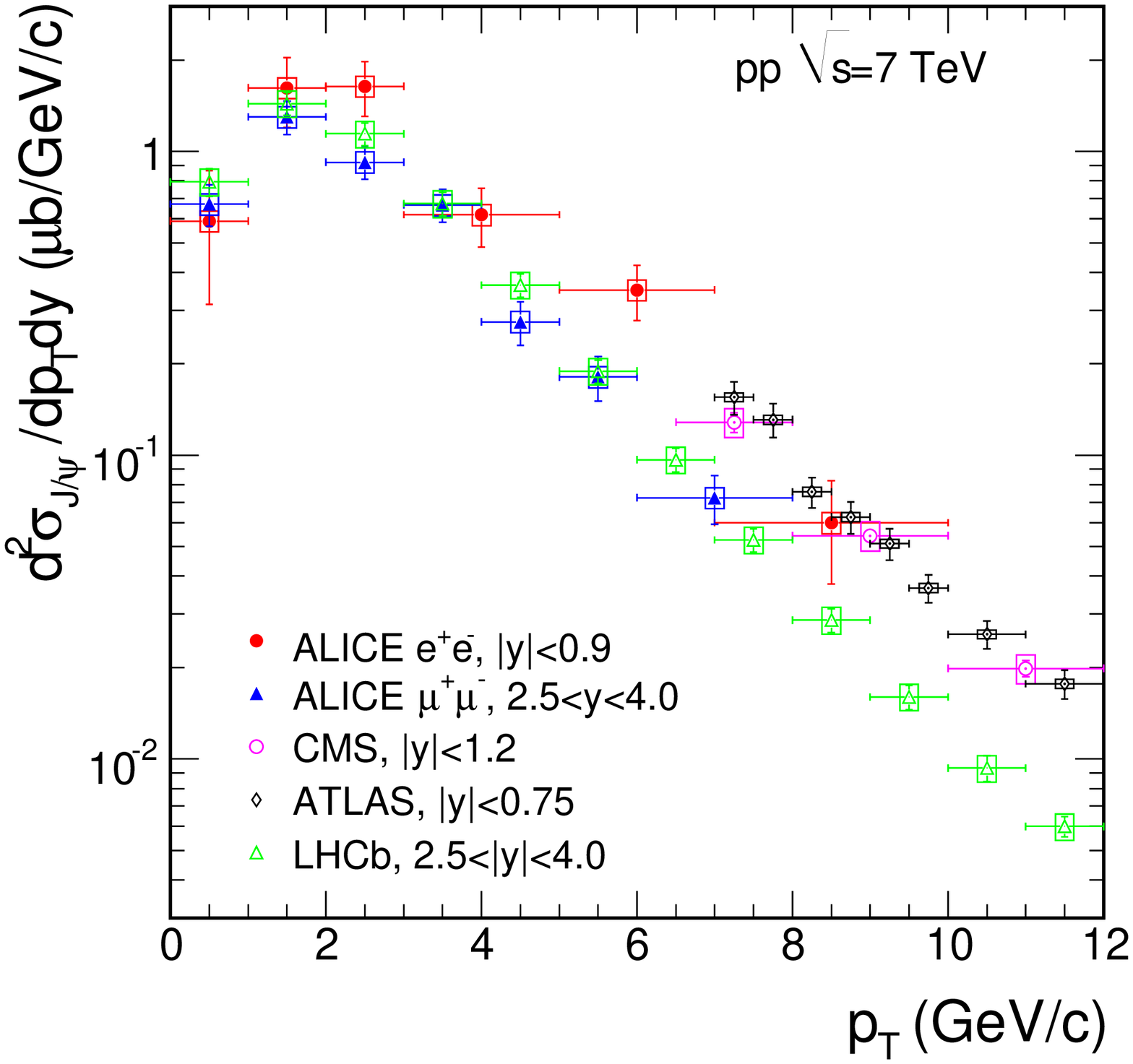}
	   \includegraphics[width=0.45\textwidth,height=0.23\textheight]{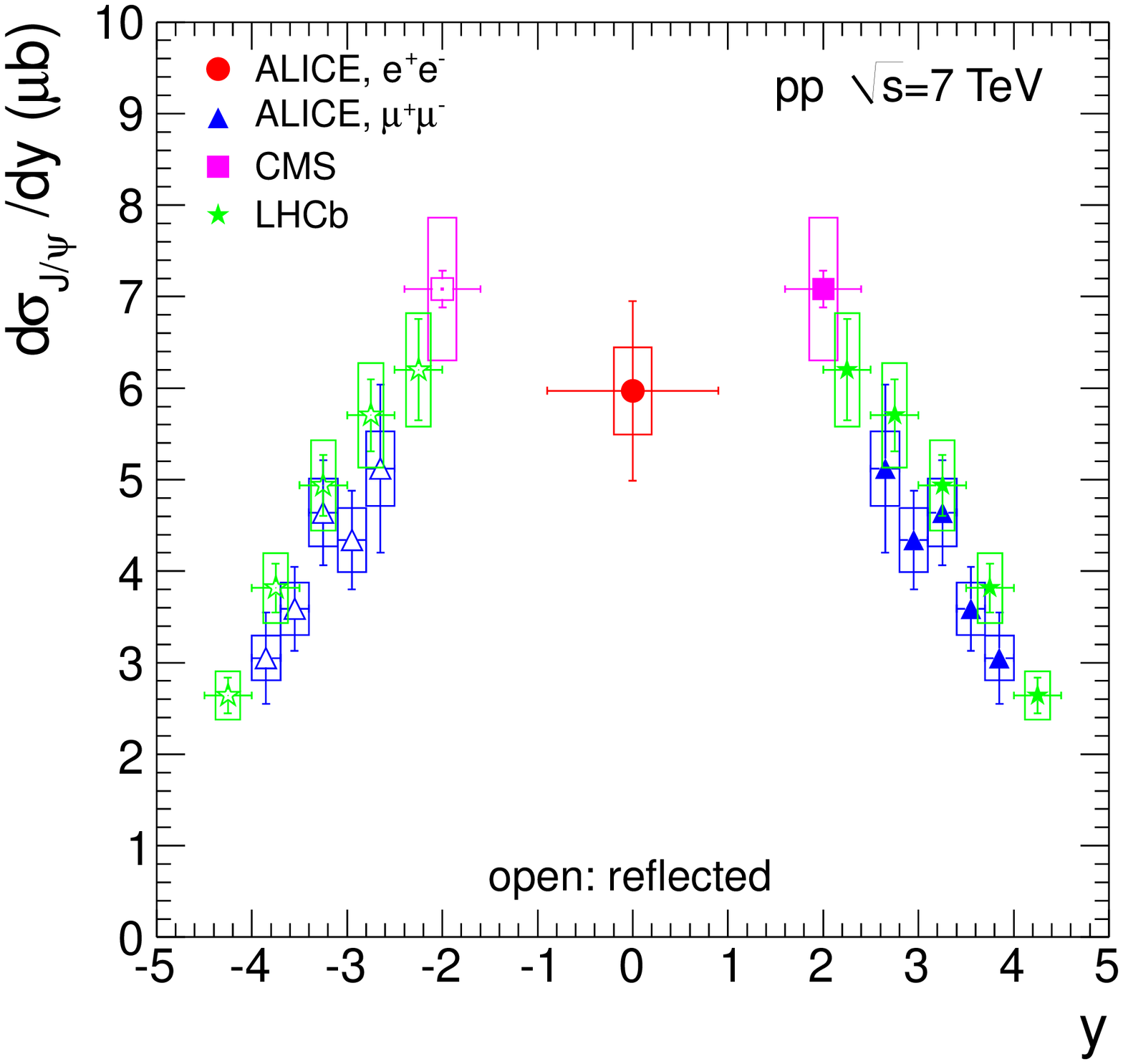}
           \label{fig:pT}
   \caption{Left: $\mathrm{d}^2\sigma_{\rm J/\psi}/\mathrm{d}p_{\rm T}\mathrm{d}y$ for
the midrapidity range and for the forward rapidity data.  Right: $\mathrm{d}\sigma_{\rm J/\psi}/\mathrm{d}y$. Both distributions are compared with results from the other LHC experiments~\cite{Kha10,Aai11,Aad11}. The error bars represent the quadratic sum of the statistical and systematic errors, while the systematic
uncertainties on luminosity are shown as boxes. From~\cite{paper}.}
\end{figure}

The ALICE results on J/$\psi$ at mid-rapidity are based on a data sample corresponding to an integrated luminosity of  3.9~nb$^{-1}$. Particle identification is performed via energy loss in the TPC; tracking uses both ITS and TPC. A hit in one of the two SPD layers (3.9 and 7.6~cm in radius) is required, in order to minimise the effect of $\gamma$ conversions. Cuts on the electron $p_{\rm T}$ (> 1~GeV/c) and on the track quality ($\chi^2$, number of TPC clusters) are applied. The J/$\psi$ yield is extracted by subtracting from the di-electron invariant mass spectrum the corresponding like-sign background distribution (Fig.~\ref{fig:signal}). The efficiency correction is performed via detailed simulation. 
The measured integrated J/$\psi$ cross section in |y|<0.9, based on a sample of $\simeq$250~J/$\psi$ is: \hbox{$10.7 \pm 1.2 (stat.) \pm 1.7 (syst.) ^{+1.6}_{-2.3} (syst. pol)~\mu b$}. Besides polarisation, the main contributions to the systematic uncertainty arise from efficiency corrections (11\%), signal extraction (8.5\%) and luminosity normalisation (8\%).  The $p_{\rm T}$-differential cross section down to $p_{\rm T}$=0 in |y|<0.9 has also been measured (Fig.~\ref{fig:pT}).

\section{J/$\psi$ $\rightarrow$ ${\rm \mu^+\mu^-}$ at forward rapidity}\label{sec:jpsimumu}

The ALICE results on J/$\psi$ production at forward-rapidity in the muon spectrometer are based on a data sample corresponding to an integrated luminosity of 15.6 nb$^{-1}$. The event selection requires at least one reconstructed vertex in the SPD and position matching between the trigger and tracking information for at least one of the reconstructed muon tracks. A cut on the radial coordinate at the exit of the front absorber is also performed, in order to reject muons emitted at small angle, that have travelled a significant distance in the thick beam shield. The invariant mass spectrum is fitted with a Crystal Ball function\cite{cb} and a double exponential for the background (Fig.~\ref{fig:signal}). Trigger and track reconstruction efficiency are evaluated via detailed simulation. Besides polarisation, the main contributions to the systematic uncertainty arise from luminosity normalisation (8\%) and signal extraction (7.5\%). The measured integrated cross section in \hbox{2.5<y<4}, based on a sample of~$\simeq$1900~J/$\psi$, is: \hbox{$6.31 \pm 0.25 (stat.) \pm 0.80 (syst.) ^{+0.95}_{-1.96} (syst. pol)~\mu b$}. The results from ALICE and LHCb~\cite{Aai11} in the common rapidity window are in good agreement. The $p_{\rm T}$ and rapidity differential cross sections have also been measured (Fig.~\ref{fig:pT}).

%\begin{figure}[!hbtp]
%   \centering
%           \includegraphics[width=0.45\textwidth,height=0.2\textheight]{2011-May-05-fig3.eps}
           
%   \caption{Invariant mass distribution for opposite-sign muon pairs.  The fitted J/$\psi$ and $\psi$ (2S) contributions, as well as the background, are also shown.}

%\end{figure}

%\begin{figure}[!hbtp]
%  \centering
%          \includegraphics[width=0.45\textwidth,height=0.23\textheight]{jpsi_dsdy1_SPD0.eps}
%    \label{fig:rapDistr}       
%   \caption{$\mathrm{d}\sigma_{\rm J/\psi}/\mathrm{d}y$ (from~\cite{paper}), compared with results
%from the other LHC experiments~\cite{Kha10,Aai11,Aad11}.
%The error bars represent the quadratic sum of the statistical and systematic
%errors, while the systematic uncertainties on luminosity are shown as boxes.}
%\end{figure}

\section{Conclusions and prospects for $Pb-Pb$ collisions}

The ALICE experiment has measured the inclusive integrated and differential (in transverse momentum and rapidity) J/$\psi$ cross sections in p-p collisions at $\sqrt{s}$=7~TeV, in a broad rapidity range and down to $p_{\rm T}$=0. The ALICE results and those of the other LHC experiments are in good agreement in the common kinematical regions.

\begin{figure}[!hbtp]
   \centering
           \includegraphics[width=0.45\textwidth,height=0.23\textheight]{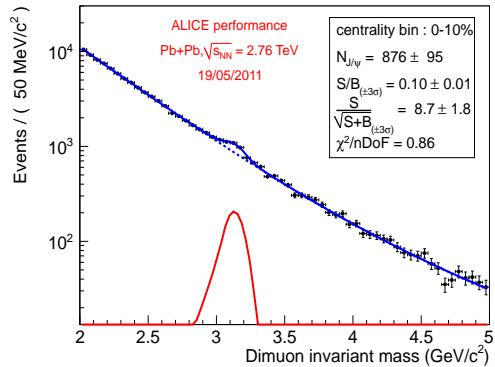}
     \label{fig:PbPb}      
   \caption{Invariant mass distribution for opposite-sign muon pairs in Pb-Pb collisions, in the centrality bin 0-10\%.  The fitted J/$\psi$ and background contributions are also shown.}

\end{figure}

In 2010, the LHC delivered the first data sample in Pb-Pb collisions at $\sqrt{s_{NN}}$=2.76~TeV. In one month of data-taking, a statistically signficant J/$\psi$ sample has been collected (Fig.~\ref{fig:PbPb}): the analysis is ongoing and first results have been obtained~\cite{qm}. 

%%%%%%%%%%%%%%%%%%%%%%%%%%%%%%%%%%%%%%%%%%%%
%% Sample figure:
%%
%% The option [height=...] scales the picture to the given height,
%% without it it would be printed at its nominal size
%%%%%%%%%%%%%%%%%%%%%%%%%%%%%%%%%%%%%%%%%%%%

%%%%%%%%%%%%%%%%%%%%%%%%%%%%%%%%%%%%%%%%%%%%
%% SAMPLE TABLE
%%
%% Shows the use of \tablehead and \tablenote
%% macros
%%%%%%%%%%%%%%%%%%%%%%%%%%%%%%%%%%%%%%%%%%%%

%%%%%%%%%%%%%%%%%%%%%%%%%%%%%%%%%%%%%%%%%%%%%%%%
%% BACKMATTER
%%%%%%%%%%%%%%%%%%%%%%%%%%%%%%%%%%%%%%%%%%%%%%%%

%\begin{theacknowledgments}
%\end{theacknowledgments}



\bibliographystyle{aipproc}   % if natbib is available

%\bibliography{sample}

%  \typeout{******************************************}
%  \typeout{** Please run "bibtex \jobname" to optain}
%  \typeout{** the bibliography and then re-run LaTeX}
%  \typeout{** twice to fix the references!}
%  \typeout{******************************************}

%%%%%%%%%%%%%%%%%%%%%%%%%%%%%%%%%%%%%%%%%%%
%% The following lines show an example how to produce a bibliography
%% without the help of the BibTeX program. This could be used instead
%% of the above.
%%%%%%%%%%%%%%%%%%%%%%%%%%%%%%%%%%%%%%%%%%%

%\endinput
%%
%% End of file `template-6s.tex'.
\end{document}